\documentclass[aps,prl,superscriptaddress,reprint,showpacs,floatfix]{revtex4-2}
\usepackage{graphicx}
\usepackage{dcolumn}
\usepackage{bm}
\usepackage{xcolor}
\usepackage{relsize}
\usepackage{amsmath}
\usepackage{amsfonts}
\usepackage{amssymb}
\usepackage{mathtools}
\usepackage{braket}
\usepackage{gensymb}
\usepackage{url}
\usepackage{mathrsfs}
\usepackage{enumitem}
\usepackage{physics}
\setlist[itemize]{label={}}

\usepackage[colorlinks = true,
            linkcolor = blue,
            urlcolor  = blue,
            citecolor = blue,
            anchorcolor = blue]{hyperref}
\usepackage{footmisc}
\usepackage[utf8]{inputenc}

\begin{document}

\preprint{APS/123-QED}

\title{Polalrized reservoirs in dynamics of polariton condensation}

\author{I. A. Shelykh}
\affiliation{Science Institute, University of Iceland, Dunhagi 3, IS-107 Reykjavik, Iceland}

\author{A. V. Yulin}
\affiliation{ITMO University, St. Petersburg 197101, Russia}

\begin{abstract}
We review the problem of description of the dynamics of driven-disspipative spinor polariton condensates, focusing on the terms corresponding to the coupling between a macroscopic wavefunction of the condensdate and incoherent excitonic reservoir created by a non-resonant pump. We demonstrate that the existing version of the theory breaks down in case, when reservoir has non-zero components of the Stokes vector corresponding to in-plane linear polarization. The polarization-invariant theory of reservoir to condensate coupling is formulated with use of the spin density matrix formalism.

\end{abstract}

\maketitle

\textit{Introduction.}
A system of of bosons being cooled down to sufficiently low temperature undergoes a transition into the macroscopically coherent state, known as the Bose-Einstein condensate (BEC). In this regime, the condensed fraction can be described by its macroscopical wavefunction, also known as an order parameter of the system, which is an ensemble average of the field operator $\psi(\mathbf{r},t)=\langle \hat{\psi}(\mathbf{r},t)\rangle$. The simplest version of mathematical formalism for the description of time evolution of a macroscopic wavefunction consists of using a mean-field approximation, which for the simplest case of scalar bosons results in the famous Gross-Pitaevskii equation \cite{PitaevskiiBook}:
\begin{align}
    i\hbar\frac{\partial\psi}{\partial t} =-\frac{\hbar^2}{2m}\nabla^2\psi+U\psi+g|\psi|^2\psi,\label{GPEq}
\end{align}
where \(m\) is the mass of the particles, \(U(\mathbf{r})\) is the external trapping potential, and \(g\) is the interaction coefficient proportional to the \(s\)-wave scattering length. The Gross-Pitaevskii equation is routinely used for the description of dynamics of condensates of cold atoms and despite its remarkable simplicity usually gives a good fit of the experimental data. 

The characteristic temperature of the BEC is inversely proportional to the mass of the particles, and for atomic condensates it lies in the nano-Kelvin regime, thus hindering perspectives of their practical implementation. This stimulated the search of analogs of the BEC in solid state systems where the effective masses of bosonic elementary excitations are orders of magnitude smaller than for cold atoms. Indirect excitons in double quantum wells \cite{Butov2002}, magnons \cite{Borisenko2020}, cavity polaritons \cite{Kasprzak2006,Balili2007} and even photons in nonlinear media \cite{KLaers2010} were proposed as candidates for the achievement of high-temperature BEC. 

In the present paper, we focus on the case of cavity polaritons, where record high critical temperatures of BEC were reported \cite{Christopoulos2007}. Polaritons, hybrid elementary excitoations in quantum microcavities in the regime of strong light-matter coupling, possess extremely small effective mass (about $10^{-5}$ of the mass of free electrons) and macroscopic coherence length (in the mm scale) \cite{Ballarini2017}. Moreover, they interact with each other and other particles in the system, including phonons and free electrons, opening the way to efficient thermalization in polariton ensembles \cite{Malpuech2002,Richard2005,Sun2017}.

Polariton systems have certain important peculiarities that distinguish them from cold atoms. First of all, they possess a finite lifetime (up to hundreds of picoseconds in the best quality samples) \cite{Mukherjee2021} and can be created by the external pump. This makes polariton BECs inherently dynamic, unlike atomic ones. Although the formation of a quasi-thermal polariton distribution has been routinely observed \cite{Richard2005,Kasprzak2007,Sun2017}, the condensation process is not driven by thermalization alone, but by its delicate interplay with the pump and decay processes. The role of the coupling of a condensate to an incoherent excitonic reservoir therefore becomes crucial \cite{Galbiati2012,Estrecho2018}.

Second, polaritons possess a peculiar spin (or pseudospin) structure \cite{ShelykhReview}. Similarly to photons, they have two possible spin projections on the structure growth axis corresponding to the two opposite circular polarizations, which makes polariton spin dynamics directly accessible experimentally via optical polarization measurements. Importantly, nonlinear polaritonic response is strongly polarization-dependent, as polaritons of the same circular polarization interact orders of magnitude stronger than polaritons with opposite circular polarizations \cite{Glazov2009}, and any coherent theory of the polariton BEC should take into account the spinor nature of polaritons. 

The effects of incoherent pumping and the coupling between scalar polaritons and incoherent excitons are generally described within the framework of the Wouters--Carusotto model \cite{WoutersCarusotto2007}, which often provides satisfactory agreement with existing experimental data. Straightforward generalization of this theory for the spinor case \cite{Borgh2010} also exists. However, as we will show below, it contains certain important drawbacks. In particular, the condensate to reservoir coupling term in it is not polarization-invariant, and can give wrong results in the case when linear polarization degree of the reservoir is different from zero. 

\textit{Theoretical Model.}
In the existing version of the model, the system of polarized polaritons is described by the two-component macroscopic wavefunction of the condensate $\Psi(\mathbf{r},t)=(\psi_+(\mathbf{r},t);\psi_-(\mathbf{r},t))^T$, with indices $\pm$ corresponding to the two circular polarizations, and the reservoir, characterized by concentrations of its circular polarized components $n_{R\pm}(\mathbf{r},t)$. In the simplest case when all types of polarization splittings in the condensate can be neglected, there is no external trapping potential, and mobility of incoherent reservoir can be neglected, the couples system of the dynamic equations for the polariton ensemble reads:
\begin{align}
i \frac{\partial \psi_\pm}{\partial t}
&=
\left[\frac{i}{2}\left(Rn_{R\pm}-\gamma\right)
-\frac{\hbar }{2m_{LP}}\nabla^2+g|\psi_\pm|^2+2\tilde g\,n_{R\pm}
\right]\psi_\pm \label{Eq:psipm}
\\
\frac{\partial n_{R_\pm}}{\partial t}
&=-(R|\psi_\pm|^2+\gamma_R)n_{R\pm}+P_\pm(\mathbf r),
\label{eq:S1_reservoir_full}
\end{align}
where the parameter $R$ is the reservoir-condensate scattering rate, $\gamma,\gamma_R$ correspond to the finite lifetime of the condensate and the reservoir, the parameter $g$ describes the interaction strength within the condensate, $\tilde{g}$ the strength of the interaction between condensed polaritons and reservoir excitons (in both cases we only account for the interaction between the same circular polarization components \cite{Ciuti1998,Glazov2009}), $P_\pm$ is a polarization-dependent incoherent pump. As our focus here will be on coupling between polarions to incoherent excitons only, to make things more transparent, let us leave only the corresponding term in the above equations:
\begin{equation}
\frac{\partial \psi_\pm}{\partial t}=\frac{R}{2}n_{R\pm}\psi_\pm, \frac{\partial n_{R_\pm}}{\partial t}=-R|\psi_\pm|^2n_{R\pm}.
\label{eq:S1_reservoir_simpl}
\end{equation}
Rewritten in terms of the concentrations of the condensed components $n_\pm=|\psi_\pm|^2$ this gives:
\begin{equation}
 \frac{\partial n_{\pm}}{\partial t}=-\frac{\partial n_{R_\pm}}{\partial t}=-Rn_\pm n_{R\pm}.    
\end{equation}
This equation has clear physical meaning: the scattering rate from the reservoir to the condensate is proportional to the product of their concentrations, which corresponds to the stimulated nature of the scattering, the total number of the particles, as well as their circular polarization is conserved, and all is good so far. 

However, let us now change the polarization basis, going from circular polarized components $\psi_\pm$ to linear polarized ones, 
\begin{equation}
    \psi_x=\frac{1}{\sqrt{2}}\left(\psi_++\psi_-\right), \psi_y=\frac{1}{\sqrt{2}}\left(\psi_+-\psi_-\right).
\end{equation}
The dynamic equation for the concentration of linear x-polarized polaritons $n_{x,y}=|\psi_{x,y}|^2$ now reads:
\begin{widetext}
\begin{equation}
 \frac{\partial n_{x,y}}{\partial t}=R\left[n_+n_{R+}+n_-n_{R-}\pm\frac{1}{2}\left(n_{R+}+n_{R-}\right)\left(\psi_+^\ast\psi_-+\psi_-^\ast\psi_+\right)\right]\neq Rn_{x,y}n_{Rx,y}.    
\end{equation}
\end{widetext}

The system of equations \ref{eq:S1_reservoir_simpl} thus contains a fundamental difficulty: it is not covariant under change of the polarization basis. 

To better understand the origin of the problem, let us rewrite Eqs.\ref{eq:S1_reservoir_simpl} in the following form:
\begin{align}
\frac{\partial}{\partial t} \begin{pmatrix} \psi_+ \\ \psi_-\end{pmatrix}&=\frac{R}{2}\begin{pmatrix} n_{R+} && 0 \\ 0 && n_{R-} \end{pmatrix}\begin{pmatrix} \psi_+ \\ \psi_-\end{pmatrix},\\
\frac{\partial}{\partial t}\begin{pmatrix} n_{R+} && 0 \\ 0 && n_{R-} \end{pmatrix} &=-R\begin{pmatrix} n_{R+} && 0 \\ 0 && n_{R-} \end{pmatrix}\begin{pmatrix} |\psi_+|2 && 0 \\ 0 && |\psi_-|^2 \end{pmatrix}.
\end{align}
One can note the following:

1. The system of equations above is not invariant under the unitary transformation describing the change of the polarization basis, $\Psi'=\hat{U}\psi,\hat{U}^\dagger=\hat{U}^{-1}$, except for the simplest case of a non-polarized reservoir, $n_{R+}=n_{R-}=n_R/2$.

2. The matrices $\mathrm{diag}[n_{R+};n_{R-}]$ and $\mathrm{diag}[|\psi_{+}|^2;|\psi_{-}|^2]=\mathrm{diag}[n_{+};n_{-}]$ resemble spin density matrices of the reservoir and condensate in the circular, but only for the case when linear polarized components, defining the off-diagonal matrix elements, are absent.

This latter observation allows us to write the correct dynamic equations replacing the defect spin density matrices by correct ones, which gives:
\begin{align}
\frac{\partial \Psi}{\partial t}&=\frac{R}{2}\hat{\rho}_R\Psi, \label{Eq:Psi}\\
\frac{\partial\hat{\rho}_R}{\partial t} &=-\frac{R}{2}\left(\hat{\rho}_R\hat{\rho}+\hat{\rho}\hat{\rho}_R\right), \label{Eq:rhoR}
\end{align}
where in circular poalrization basis the density matrices of the condensate $\hat{\rho}$ and reservoir $\hat{\rho}$ read:
\begin{align}
\hat{\rho}=\begin{pmatrix} |\psi_+|2 && \psi_+^\ast\psi_- \\ \psi_-^\ast\psi_+ && |\psi_-|^2 \end{pmatrix}, \hat{\rho}_R=\begin{pmatrix} n_{R+} && \rho_{12} \\ \rho_{12}^\ast && n_{R-} \end{pmatrix}.\label{Eq:DensMatr} 
\end{align}
Note that Eq.\ref{Eq:DensMatr} is invariant under the unitary transformation defining the change of polarization basis:
$\Psi'=\hat{U}\psi, \hat{\rho}'=\hat{U}\hat{\rho}\hat{U}^\dagger, \hat{\rho}_R'=\hat{U}\hat{\rho}_R\hat{U}^\dagger, \hat{U}^\dagger=\hat{U}^{-1}$.

Decomposing the spin density matrices of the condensate and reservoir by the basis consisting on the unity matrix $I$ and set of Pauli matrices $\sigma_j$,
\begin{align}
    \hat{\rho}=\frac{1}{2}\left(nI+\mathbf{S}\cdot\boldsymbol{\sigma}\right), \hat{\rho}_R=\frac{1}{2}\left(n_RI+\mathbf{S}_R\cdot\boldsymbol{\sigma}\right)
\end{align}
one gets the following dynamic equations for the occupancies and pseudospin vectors of condensate and reservoir:
\begin{align}
    \frac{\partial n}{\partial t}=-\frac{\partial n}{\partial t}=2R\left(nn_R+\mathbf{S}\cdot\mathbf{S}_R\right), \label{Eq:n} \\
    \frac{\partial \mathbf{S}}{\partial t}=-\frac{\partial \mathbf{S}_R}{\partial t}=2R\left(n_R\mathbf{S}+n\mathbf{S}_R\right), \label{Eq:S}
\end{align}
which coincide with semiclassical Boltzmann equations for spinor polaritons \cite{KKavokin2004}.

Combining Eqs.\ref{Eq:psipm} and \ref{eq:S1_reservoir_full} with Eqs.\ref{Eq:Psi} and \ref{Eq:rhoR} one gets the following system of dynamic equations, which is the main result of the current paper:
\begin{widetext}
\begin{align}
i \frac{\partial \Psi}{\partial t}
&=
\left[\frac{i}{2}\left(R\hat{\rho}_R-\gamma\right)
-\frac{\hbar }{2m_{LP}}\nabla^2+\begin{pmatrix} g|\psi_+|^2+\tilde g\,n_+ && 0 \\ 0 && g|\psi_-|^2+\tilde g\,n_-\end{pmatrix}
\right]\Psi \label{Eq:PsiGen}
\\
\frac{\partial \hat{\rho}_R}{\partial t}
&=\frac{1}{2}\left(PI+\mathbf{S}_P\cdot\boldsymbol{\sigma}\right)-\Gamma\rho_R-\frac{R}{2}\left(\hat{\rho}_R\hat{\rho}+\hat{\rho}\hat{\rho}_R\right)+\begin{pmatrix} \frac{n_{R-}-n_{R+}}{\tau_\perp} && -\frac{\rho_{12}}{\tau_\parallel} \\ -\frac{\rho_{12}^\ast}{\tau_\parallel} && \frac{n_{R+}-n_{R-}}{\tau_\perp}\end{pmatrix},
\label{eq:ResGen}
\end{align}
\end{widetext}
where the density matrices of the condensate and reservoir are defined as in Eq.\ref{Eq:DensMatr}, $P$ and $\mathbf{S}_P$ are the intensity of the incoherent pump and its Stokes vector, $\rho_{12}=S_{Rx}-iS_{Ry}$ corresponds to the linear polarization of the reservoir. 

The last term in Eq.\ref{eq:ResGen} describes spin relaxation inside the reservoir, with $\tau_\perp$ and $\tau_\parallel$ being characteristic relaxation times of circular and linear polarizations. Note that in  general these two times are different and can depend on reservoir concentration due to, e.g., a fast loss of linear polarization in the reservoir due to spontaneous exciton-exciton scattering \cite{Shelykh_2009}. Note, that in the limit of very fast in-plane spin relaxation, $\tau_\parallel\rightarrow 0$ the off-diagonal terms of the density matrix of a reservoir roll to zero, and our equations coincide with those used previously.

Let us now discuss an example illustrating the differences between cases where the component $\rho_{12}$ can be neglected and where it significantly affects the polarization dynamics of the condensate. For this purpose, we consider a simplified scenario in which spatial effects are negligible, i.e., the sample is incoherently excited by a light beam with a large aperture. Consequently, we neglect the $\nabla^2 \phi$ term in Eq. (\ref{Eq:PsiGen}).

The system is excited incoherently by elliptically polarized light. We focus on the case where the polarization ellipse is strongly elongated along the $x$-axis, such that $S_x \gg S_z$ and $S_y=0$. Experiments have shown that under these conditions, the condensate formed under the pump is quasi-linearly polarized, and its polarization plane rotates with an angular velocity determined by the pump ellipticity  \cite{gnusov2020optical,baryshev2022engineering}. We now investigate how the polarization dynamics depend on the relaxation time $\tau_{\parallel}$. In the limit $\tau_{\parallel} \rightarrow 0$, our model reduces to the well-known case of two condensate polarizations interacting with two reservoirs of opposite circular polarizations \cite{carusotto2013quantum}. However, for longer relaxation times, a clear qualitative difference emerges.

For the numerical simulations, we use typical parameters. The results do not depend qualitatively on the exact parameter values; the simulations presented here are performed using parameters taken from  \cite{dall2014creation}: $\gamma=0.1$ps$^{-1}$, $\Gamma=0.35$ ps$^{-1}$, $R=0.01$ ps$^{-1}$$\mu$m$^2$, $g=0.0091$ps$^{-1}$$\mu$m$^2$, $\tilde g=0.0182$ps$^{-1}$$\mu$m$^2$. We set the relaxation time $\tau_{\perp}=3$ ps and note that the condensate dynamics remain very similar for longer $\tau_{\perp}$.

We first consider the case of a relatively long relaxation time, $\tau_{\parallel} = 0.1$ ps. The initial conditions for the condensate order parameter $\psi$ are taken as random values of small magnitude. We monitor the formation of the condensate under the pump and examine how the final state depends on the pump ellipticity. The pump power is set to approximately twice the threshold value, with fixed components $S_{Px}=30$, $S_{Py}=0$. We then vary $S_{Pz}$ and study the dependence of the final condensate state on this parameter.

\begin{figure}[t]
\includegraphics [width=1\columnwidth]{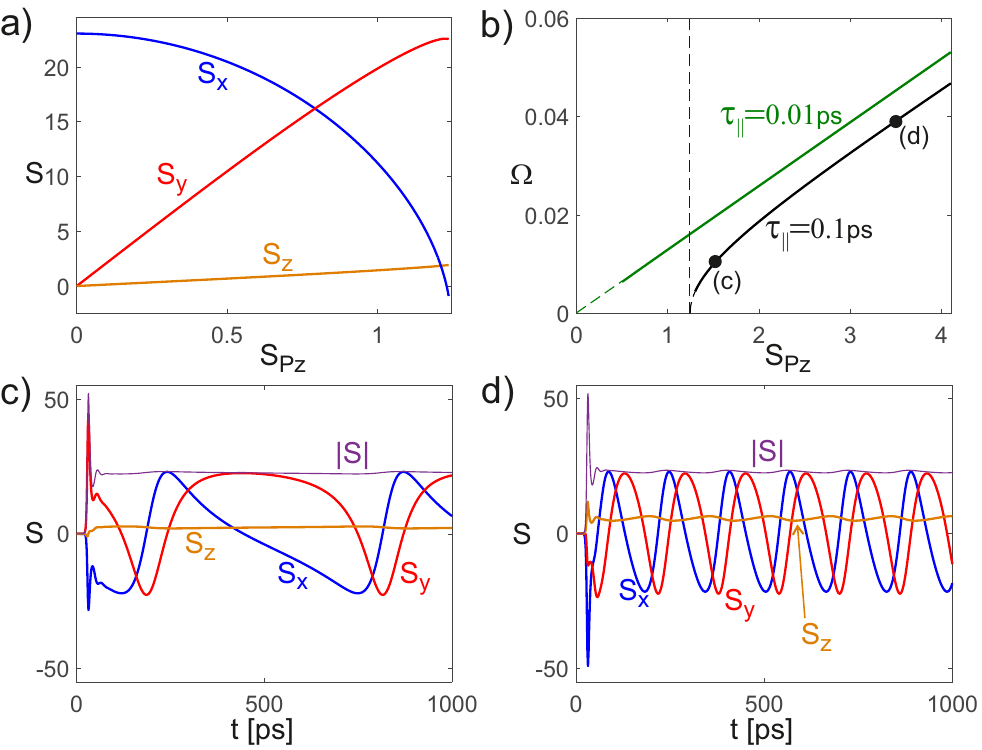}
\centering
\caption{(a) The dependencies of the Stokes components of the stationary time-independent state on $z$ component of the Stokes vector $S_{Pz}$ of the pump. The blue curve corresponds to $S_x$, red curve - to $S_y$ and orange curve - to $S_z$ component of the Stokes vector of the condensate. The relaxation time .$\tau_{\parallel}=0.1$ ps$^{-1}$. (b) The dependency of the poarization plane angular velocity (psudo-spin precession frequency) as a function of $z$ component of the Stokes vector of the pump. The thicker dark curve is for $\tau_{\parallel}=0.1$ ps$^{-1}$, the thinner green line is for $\tau_{\parallel}=0.01$ ps$^{-1}$. The dashed curve shows the square root fit for the numerically calculated dependency. The threshold value of $S_{Pz}$ for $\tau_{\parallel}=0.1$ ps$^{-1}$ is shown by vertical dashed line. (c),(d) The temporal evolution of the Stokes components calculated for the condensate excited by the pump with $S_{Pz}=1.5$ and $P=3.5$ correspondingly. The relaxation time $\tau_{\parallel}=0.1$ ps$^{-1}$. The blue curve is for $S_x$, red curve - for $S_y$ and orange curve - for $S_z$, the length of Stokes vector is shown by the magenta curve. See text for more detail. }

\label{fig1}
\end{figure}

To characterize the condensate, we introduce its Stokes parameters: $S_x=2 \Re \left( \psi_{+} \psi_{-}^{*} \right)$, $S_y=-2 \Im \left( \psi_{+} \psi_{-}^{*} \right)$  and $S_z=| \psi_{+}|^2-| \psi_{-}|^2$, with the total intensity $S=\sqrt{S_x^2+S_y^2+S_z^2}$. The simulations reveal the existence of a threshold value of $S_{Pz}$ such that below this value, the final state of the condensate is time-independent. The dependence of the Stokes vector components of the stationary condensate on $S_{Pz}$ is shown in Fig.~\ref{fig1}(a).

As expected, for a purely linearly polarized pump ($S_{Pz}=0$), the condensate is polarized along the $x$-axis, with $S_{y,z}=0$. As the pump ellipticity increases, the orientation of the condensate polarization plane changes. Near the threshold value of $S_{Pz}$, the polarization plane is oriented at an angle close to $\frac{\pi}{4}$ relative to the $x$-axis. Notably, the ellipticity of the condensate remains relatively small, with $\frac{S_z}{S} < 0.1$.

When the pump ellipticity exceeds the threshold value, the polarization plane of the condensate begins to rotate, with an angular velocity that increases with $S_{Pz}$ (see Fig.~\ref{fig1}(b)). The temporal evolution of the condensate Stokes parameters is shown in Fig.~\ref{fig1}(c) and (d) for $S_{Pz}=1.5$ and $S_{Pz}=3.5$, respectively. Close to the threshold, the time dependencies $S_x(t)$ and $S_y(t)$ are distinctly nonsinusoidal, indicating nonuniform rotation of the polarization plane. For larger values of $S_{Pz}$, the rotation frequency increases and the motion becomes smoother. Additionally, the rotation induces oscillations in the $z$ component and in the total length of $\vec S$.

If we significantly decrease the relaxation time $\tau_{\parallel}$, the threshold value of $S_{Pz}$ becomes very small, and the dependence of the rotation frequency on $S_{Pz}$ becomes nearly linear (see Fig.~\ref{fig1}(b)). In this regime, the evolution of the Stokes vector becomes essentially sinusoidal. We therefore conclude that the pump polarization creates an effective potential that pins the orientation of the condensate polarization plane. A short relaxation time of the off-diagonal components of the density matrix $\rho$ wipes out the memory of the condensed polaritons regarding the direction of the reservoir polarization; nevertheless, the ellipticity of the reservoir polarization remains an important factor.

In summary, we presented a theory of the dynamical polariton condensation that accounts for the possibility of reservoir polarization. The theory possesses the required property of the invariance under change of polarization basis.

\bibliography{main}   

\end{document}